# Refractive multi-conjugate adaptive optics for wide-field atmospheric turbulence correction

*Tommaso Furieri[1], Stefano Bonora[1,*]*

[1]*Consiglio Nazionale delle Ricerche, Institute of Photonics and Nanotechnology, via Trasea 7, 35131, Padova, Italy*

*\*stefano.bonora@cnr.it*

**Abstract:** Multi-Conjugate Adaptive Optics (MCAO) is essential for increasing the corrected Field-of-View (FoV) in astronomical imaging and potentially for free-space optical communications, particularly for small-aperture, transportable systems. We demonstrate the viability and performance of a Refractive-MCAO system utilizing a novel multi-actuator Deformable Lens (DL) as the wavefront correction element. Unlike conventional Deformable Mirrors (DMs), the transmissive nature of the DL simplifies the optical train, making it ideal for compact setups. Using a Shack-Hartmann Wavefront Sensor (SH-WFS) in conjunction with two DLs conjugated to different atmospheric layers, we achieved an extension of the isoplanatic patch up to three times the uncorrected atmospheric isoplanatic angle under moderate turbulence $D/r_0 = 2$.

We tested the MCAO system in a setup that emulates a free space optical communication for compact transportable system. In a double-channel, single-mode fiber coupling experiment we demonstrated the efficiency of this method.

## 1. Introduction

Adaptive optics (AO) techniques enable real-time correction of optical wavefront distortions induced by spatial and temporal variations in the refractive index of the propagation medium. Originally developed for astronomical imaging through atmospheric turbulence [1], AO has found widespread application in microscopy [2], ophthalmology [3], and high-power laser systems [4]. In its classical implementation, AO employs a single deformable mirror (DM) conjugated to the pupil plane of the imaging system to compensate for phase aberrations measured along a single point in the field of view (called SCAO Single Conjugated Adaptive Optics). However, because optical turbulence or sample-induced aberrations are generally distributed along the line of sight, single-conjugate correction is fundamentally limited in field of view and fails to provide uniform image quality across extended scenes [5].

To address this limitation, multi-conjugate adaptive optics (MCAO) introduces multiple deformable mirrors, each optically conjugated to a distinct layer within the aberrating volume [6]. This architecture enables partial reconstruction of the three-dimensional (3D) structure of the refractive index perturbations and improves correction uniformity over larger fields. The underlying principle of MCAO can be interpreted as a discrete approximation to the ideal, continuous optical transformation that would exactly invert the volumetric aberration structure. Theoretically, if one had access to an infinite number of adaptive optical elements, each capable of applying a spatially varying phase shift at an arbitrary depth, the cumulative effect would reproduce in the image space the exact conjugate of the aberration distribution in the object space. The result would be perfect aberration cancellation, yielding diffraction-limited imaging over the entire field of view.

While such a continuous volumetric correction is physically unattainable with conventional reflective elements, it provides a powerful conceptual framework for alternative implementations. In particular, refractive adaptive optics, based on tunable or deformable lenses, offers an inherently transmissive means of phase modulation. Unlike reflective DMs, refractive correctors can be cascaded in series within compact optical trains, allowing for the direct approximation of distributed volumetric phase compensation through a stack of conjugate refractive layers. By modulating the local optical thickness or refractive index profile in each plane, it becomes possible to emulate the action of multiple deformable mirrors while maintaining a transmissive, alignment-friendly geometry.

In this work, we present a Refractive Multi-Conjugate Adaptive Optics (R-MCAO) architecture, which replaces conventional deformable mirrors with a series of independently controllable refractive correctors. We then describe the optical design and control strategy of the R-MCAO system, and we demonstrate its capability to restore image quality across extended fields through numerical simulation. This refractive implementation of MCAO represents a compact, and scalable alternative to conventional mirror-based systems, with potential impact on fields including astronomical instrumentation, medical and microscopy. Fig. 1 reports how R-MCAO can be implemented in a generic optical system. Considering a case where the source of the aberrations is distributed over a volume in the object space (for example in microscopy techniques that can image trough the object such as confocal, light sheet or multiphoton, or the case of atmospheric turbulence, such as in astronomy or free space laser communications) if we use a number of deformable lenses in the image space (three in the case reported in the figure), we apply a phase correction that is imaged before the object, therefore the system acts like a precorrection of the volumetric aberrations.

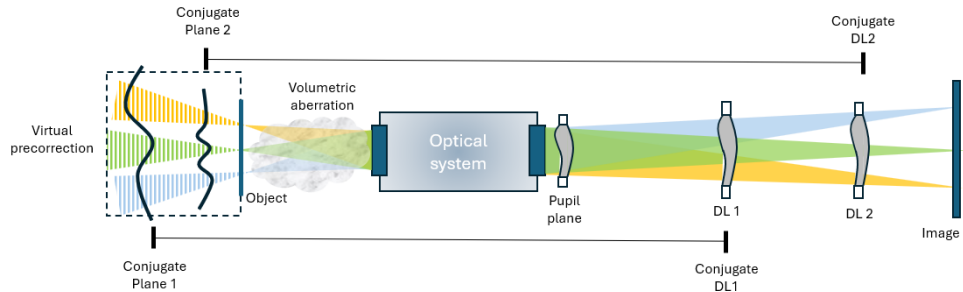

Figure 1: DL MCAO Optical Layout Simulated MCAO optical system architecture using two multi-actuator Deformable Lenses. This layout highlights the transmissive advantage of the DLs compared to a reflective DM-based setup.

A first demonstration in microscopy of R-MCAO has been reported in [8] where two MAL deformable lenses where used, exploiting the SPAM system as a feedback, over extended light sheet microscopy images. The use of MAL for R-MCAO has never been tested in the compensation of atmospheric turbulence. In this case the response time of the refractive device and of the control system has a relevant importance.

Among possible refractive correctors we used MAL because they are the only one that has at the same time high transparency (>98% between 450nm – 1200nm) and fast response time (about 1.5ms). The drawback is the limited number of actuators. This limitation will be discussed in detail in the following. Alternative refractive corrector such as optofluidic phase plate [7] and Liquid Crystals has either too limited transmission losses and slow response time.

## 2. DL-Based MCAO Simulation and Design

The DL is a multi-actuator refractive element, capable of correcting aberrations up to the $4^{th}$ Zernike order, similar to a typical DM. This capability allows for the correction of low-order modes that dominate atmospheric aberrations for small aperture telescope (up to 20cm). Several demonstrations of using the MAL in SCAO has been conducted in astronomical telescopes [9] and free space optical communications systems [10, 11] and underwater imaging [12]. To evaluate the efficiency of the use of MAL lenses in R-MCAO we carried out some simulations in an optical propagation framework and we carried out an experimental demonstration in free space optical communication system with the aim of doubling the communication capacity being able to couple the light inside two single mode fibers placed in different position in the field of view at the same time.

*Numerical Simulation Framework*

A numerical simulation framework was developed to model the volumetric atmospheric turbulence and the subsequent MCAO correction using a stack of DLs in order to evaluate its efficacy. The atmospheric modeling uses standard statistical properties of turbulence, consistent with the Kolmogorov model. The simulation focused on a small-sized telescope of 10cm aperture, relevant for imaging systems and laser communication ground to ground applications.

To parameterize the simulation with a well-known amount of turbulence we applied as turbulence profile the one obtained in vertical propagation as described in Eq. 1. The refractive index structure, shown in Eq. 1 depends on $w$, the RMS wind speed, and $A$. $A$ can be approximated with the value at ground of $C_n^2(0)$. We chose for $A$ three values that can be read as low, moderate and high turbulence conditions as specified in the following.

$$C_n^{2(z)} = 5.94 \times 10^{-23} z^{10} \frac{w}{27} e^{-z} + 2.7 \times 10^{-16} e^{-\frac{2}{3}z} + A\, e^{-10z} \qquad \text{Eq. 1}$$

The other parameters for all the simulations are shown in table 1:

**Table 1: simulation parameters.**

| Parameter | Value |
|---|---|
| **Telescope Diameter** [cm] | 10 |
| **Wind Speed** [m/s] | 25 |
| **Wavelength** [nm] | 633 |
| **Outer Scale** $L_0$ [m] | 200 |
| **Inner Scale** $l_0$ [mm] | 5 |
| **Zenith Angle** $\xi$ [rad] | 0 |
| **Atmospheric Layers** | 11 |

We then computed the relative Fried coherence length $r_0$ and isoplanatic angle $\theta_0$ (Eq. 2 and 3).

$$r_0 = \left[ 0.423\, k^2 \sec \xi \int_{path} C_n^{2(h)}\, dh \right]^{\left\{-\frac{3}{5}\right\}} \qquad \text{Eq. 2}$$

$$\theta_0 = \left[ 2.91\, k^2 \sec^{\frac{8}{3}}\xi \int_{path} C_n^{2(h)} h^{\frac{5}{3}}\, dh \right]^{\left\{-\frac{3}{5}\right\}} \qquad \text{Eq. 3}$$

All the values are shown in table 2.

Table 2: Turbulence parameters for low (A "9e-15), mid (A "5e-14) and high (A "2e-13) turbulence conditions.

| $A\ [m^{-2/3}]$ | $r_0$ [cm] | $\theta_0$ [rad] | $D/r_0$ |
|---|---|---|---|
| 9e-15 | 10.8 | 9.0e-6 | 0.93 |
| 5e-14 | 5.4 | 8.9e-6 | 1.85 |
| 2e-13 | 2.5 | 8.6e-6 | 3.94 |

The FoV angle for this analysis was chosen as 6.5 times the isoplanatic angle. A grid of 13x13 equally spaced points inside the FoV has been used for the field dependent wavefront calculation. In this way the Isoplanatic angle corresponds to a width of two points in the considered 13x13 grid (see Fig. 2). In fact it can be visually seen that the same aberration is roughly common for every 2x2 FoV points that we considered.

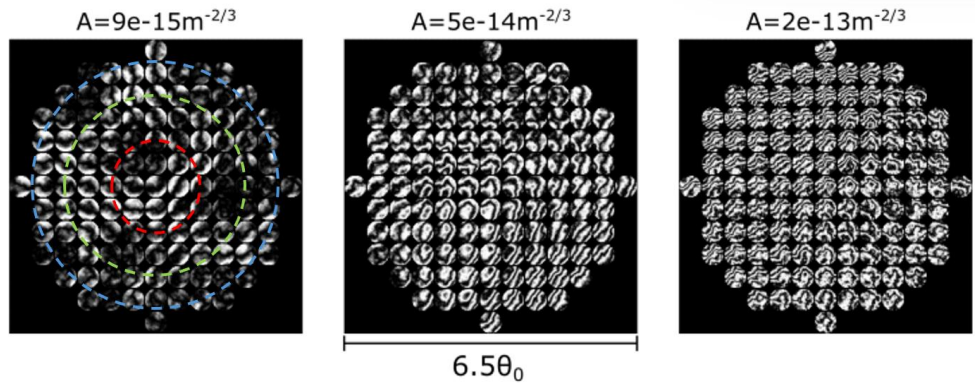

Fig. 2: phase propagation, from left to right, for low, medium and strong turbulence shown as interferograms at 633nm. The circles show the FoC: red, isoplanatic patch $\theta_0$, green "small" patch $2\theta_0$, "large" patch $3\theta_0$.

For each turbulence condition we generated 100 propagations and we tested the correction for three AO configurations:

1. pupil DL - SCAO

2. pupil DL + Conjugated plane DL1 - MCAO 1

3. pupil DL + Conjugated plane DL1 and Conjugated plane DL2 - MCAO 2

For the SCAO configuration the correction was done only in the center of the field of view, while for the MCAO ones the correction was performed in two conditions over a small and large field (Field of Correction, FoC), where for small and large we mean respectively an angle twice and three times $\theta_0$). The wavefront correction simulation was carried out from measured influence functions of MAL deformable lenses (reported in the supplementary Fig. 1 materials) with different clear aperture 10mm, 16mm and 25mm to be positioned in different position in the image space as shown in Fig. 1. The method for computing the corrected wavefront is reported in the supplementary materials. The corrected zones for the small and large patch are

1 for SCAO, 9 for the small patch and 25 for the large patch, the positions are reported in Supplementary Fig. 4.

A visual representation of a correction in case of strong turbulence can be seen in Fig. 3. In this figure each pixel of the heat maps represents the mean wavefront error (in units of $\lambda$ at 633nm) over the 100 realizations for a particular point of the FoV. We analyzed then for each of the five AO configurations the mean RMS wavefront error on both the entire FoV and only inside the targeted corrected area. The results are listed in Tables 3 and 4 where the first one shows the values considering the entire FoV while the second one considers only the FoC. In particular it emerges that in case of weak turbulence the benefit of MCAO is only marginal if compared to the implementation complexity. More interesting is the fact that if we consider the entire FoV, but the MCAO correction covers only a portion of it (FoC<FoV), the average correction is similar if not worse to the SCAO case. This is of course due to the fact that the controller only optimizes for the correcting area and usually worsen all the outer fields. It can be moreover seen that even in cases of medium turbulence (so with a $D/r_0 \approx 2$) the MCAO implementation promise to retain the almost diffraction limited correction.

Above that value MCAO shows still a significant improvement over SCAO but it is not able to fully compensate for the aberration.

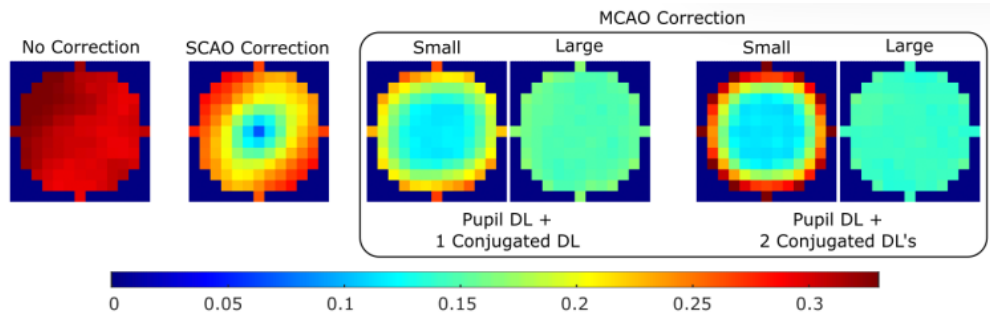

Figure 3: Simulated Isoplanatic Patch Extension in the case of strong turbulence. Comparative simulation showing the Strehl Ratio decrease versus angular distance from the guide star. The MCAO implementation significantly extends the corrected field of view, achieving an isoplanatic patch up to three times the SCAO value.

Table 3: RMS wavefront average over the entire FoV ($\lambda$ = 633nm)

| $A$ [$m^{-2/3}$] | Not Corrected | SCAO | R-MCAO1 | | R-MCAO2 | |
|---|---|---|---|---|---|---|
| | | | Small FoC | Large FoC | Small FoC | Large FoC |
| 9e-15 (Low) | 0.09$\lambda$ | 0.09$\lambda$ | 0.07$\lambda$ | 0.06$\lambda$ | 0.11$\lambda$ | 0.06$\lambda$ |
| 5e-14 (Med.) | 0.18$\lambda$ | 0.13$\lambda$ | 0.11$\lambda$ | 0.09$\lambda$ | 0.14$\lambda$ | 0.09$\lambda$ |
| 2e-13 (Strong) | 0.31$\lambda$ | 0.22$\lambda$ | 0.17$\lambda$ | 0.15$\lambda$ | 0.19$\lambda$ | 0.14$\lambda$ |

Table 4: RMS wavefront average over the corrected Field ($\lambda = 633$nm)

| A [m$^{-2/3}$] | Not Corrected | SCAO | R-MCAO1 | | R-MCAO2 | |
|---|---|---|---|---|---|---|
| | | | Small FoC | Large FoC | Small FoC | Large FoC |
| 9e-15 (Low) | 0.09λ | 0.05λ | 0.05λ | 0.06λ | 0.05λ | 0.06λ |
| 5e-14 (Med.) | 0.18λ | 0.07λ | 0.08λ | 0.09λ | 0.07λ | 0.09λ |
| 2e-13 (Strong) | 0.31λ | 0.10λ | 0.12λ | 0.15λ | 0.12λ | 0.14λ |

The simulation results confirmed the high correction capability of the DLs in an MCAO configuration. Specifically, a correction almost to the diffraction limit was observed for values of $D/r_0$ of about 2 and a good correction level for $D/r_0$ of about 4. So for an adaptive optics system designed for small aperture (10-15cm) telescopes this configuration can give important benefits with respect to configurations using SCAO.

## 3. Method

In the experimental validation we focused on a portable laser communication scenario requiring a simultaneous parallel dual communication link. The setup used a Shack-Hartmann Wavefront Sensor (SH-WFS) for real-time wavefront measurement, a necessity due to the time-varying nature of atmospheric turbulence. Two separate communication channels were simulated, necessitating independent aberration correction for each channel due to propagation angles exceeding the isoplanatic angle ($\theta_0$).

Two DLs were integrated into the optical path: one in the pupil and a second one in a position where its clear aperture is completely filled (See Fig 5 and supplementary Fig. 2). The control software, written in Python, was responsible for managing the SH-WFS measurement, computing the corrective commands via a reconstructor algorithm, and driving the DL actuators in a closed-loop with an integrative controller.

*Wavefront measurement*

In principle, for this configuration to operate correctly, the spots produced by the two beams must not overlap within the same sub-aperture of the Shack–Hartmann (SH) sensor. To achieve this, each individual lenslet area was conceptually divided into four subregions. Consequently, the centroid detection region is reduced to one quarter of its nominal size; however, this approach theoretically enables the simultaneous measurement of up to four equally spaced points within the field of view (FoV). The underlying concept is illustrated in Figure 4. Naturally, this implementation introduces a primary constraint: the angular separation between the two point sources is determined by the lenslet pitch of the SH wavefront sensor. Nevertheless, in the context of a laboratory experiment, this limitation can be mitigated through appropriate design of the optical setup.

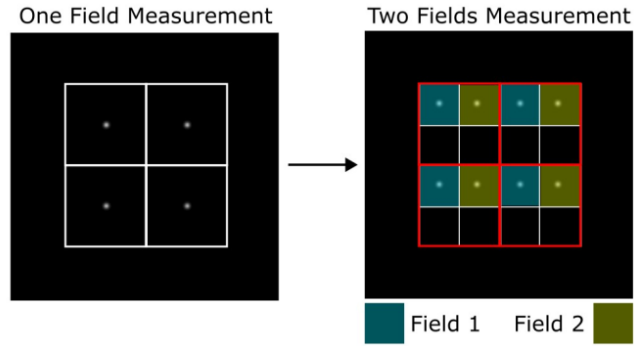

**Figure 4:** in this figure the idea of dividing each lenslet sub-aperture into four lenslets aperture is shown.

*Optical setup*

The experimental setup is shown in Figures 5. Two laser diodes (λ = 633 nm) were used as sources and combined using a 50:50 beam splitter (BS). The combined beams were collimated with a lens of focal length $f = 150$mm. The second laser source was mounted on an XYZ translation stage to allow fine adjustment of its position. Between laser source 1 and the first BS, a deformable lens (DL) was inserted to generate turbulence exclusively in the first optical channel. The output angle between the two beams was approximately 30 mrad (≈ 1.7°).

The pupil DL (aperture = 10 mm) was placed immediately after the collimating lens, while the conjugated DL (aperture = 16 mm) was positioned 200 mm downstream. A relay imaging system with a magnification ratio of 4:1 ($f_2/f_3$) was used to project the pupil plane onto a Shack–Hartmann wavefront sensor (SH-WFS, Dynamic Optics s.r.l.).

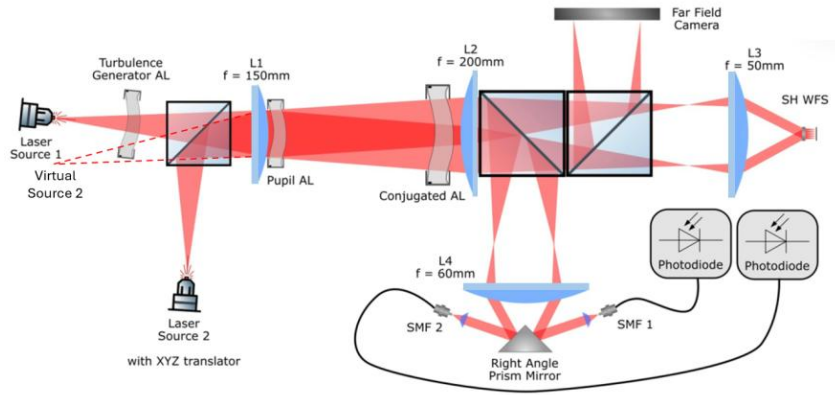

**Figure 5:** Setup used to test the free-space MCAO correction with two DLs. The angular separation of the two laser sources was 30mrad.

In the relay path, two additional beam splitters were inserted. The second BS deflected 10% of the light into a far-field camera, where the point spread functions (PSFs) of the two beams could be monitored. The first BS diverted 50% of the beam toward a lens (L3) forming an image of the pupil plane onto two single-mode fiber (SMF) couplers (Thorlabs PAF2P-15A). The beams were separated by a right-angle prism mirror before coupling. The output ends of the SMF patch cables (Thorlabs P1-630A-FC-1) were connected to two PDA36A2 switchable-gain photodiodes. The photodiode signals were acquired using an Arduino UNO microcontroller operating at a sampling rate of 750 Hz.

With this configuration, the beam diameter at the SH-WFS was 2.5 mm. An exposure time of 30 µs provided approximately 200 counts per spot peak (on a 255 count scale, just below saturation). The SH-WFS employed microlenses with a 300 µm pitch, yielding 45 measurable spots per field. The sensor operated at a maximum frame rate of 303 Hz when cropped to the area of interest.

To measure the wavefront correction capabilities of the R-MCAO system we closed the control loop keeping the second field corrected while increasing aberrations were applied to the first field sequentially for each Zernike polynomial from 1 to 9. The process was halted if any actuator reached saturation or if the mean RMS error in either field exceeded $\lambda/14$ at $\lambda = 633$ nm. The results are shown in Figure 6.

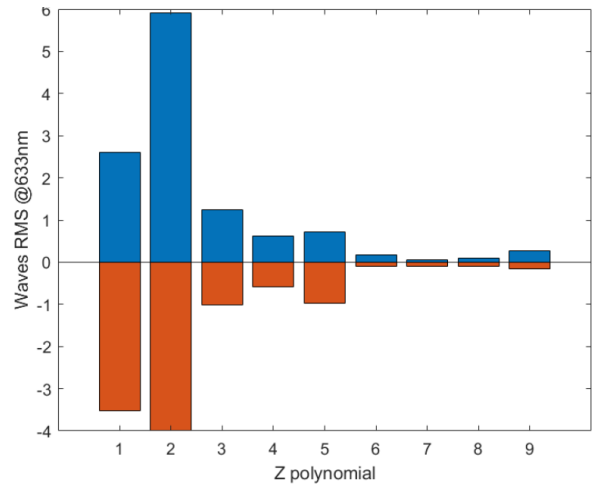

Figure 6: Measurement of the independent aberrations amplitude that this MCAO setup is able to produce.

The data exhibit the expected behavior: strong correction efficiency for low-order modes (tip, tilt, astigmatism, and defocus) and reduced response for higher-order terms such as trefoil and coma (Zernike 6–9). Visualization 1 shows the two PSFs obtained using two deformable lenses generating independent aberrations, confirming the ability of the system to perform multi-conjugate adaptive optics (MCAO) correction.

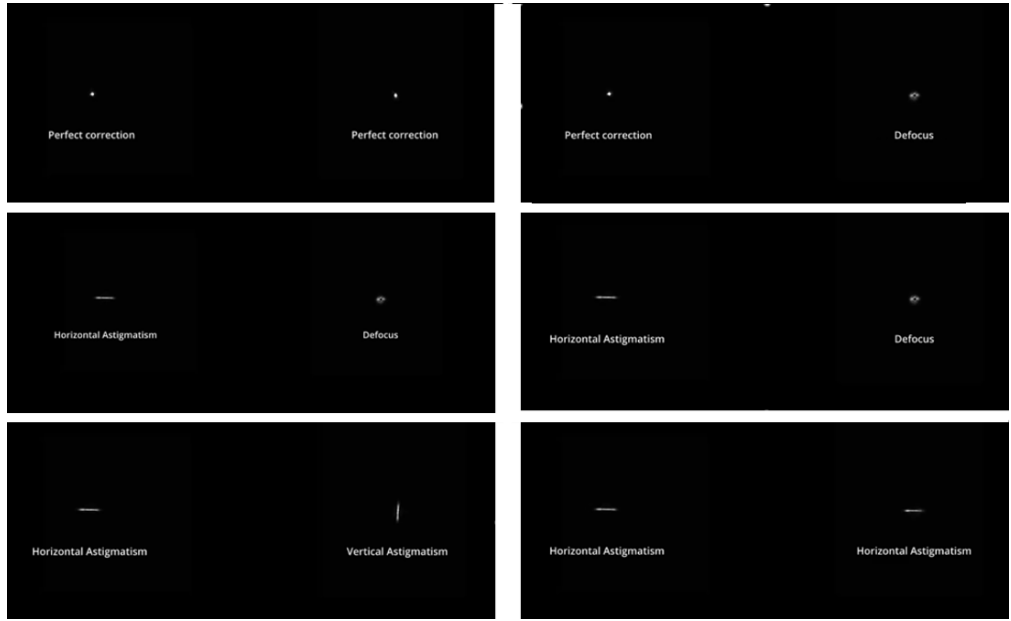

Visualization 1: examples of independent control of the two PSFs experimentally acquired by the far field camera in the two fields of the optical setup of Fig. 5. The focal spots are 14mrad separated (equivalent to 2.1mm over a focal length of 150mm).

*Turbulence simulation*

To further validate system performance, we evaluated the MCAO correction under experimentally simulated turbulence. The turbulence was introduced by inserting an additional DL in the optical path of the first field and driving it with a time series reproducing real atmospheric aberrations. Although this approach simplifies the real propagation scenario, since turbulence in the atmosphere occurs continuously rather than at a discrete plane, the absence of conjugation between the pupil DL, the conjugated DL, and the turbulence layer provides an acceptable approximation.

The turbulence time series used was recorded over a 1 km horizontal path above the sea [13]. These data were obtained with a 279 mm aperture telescope, which exceeds the 100 mm pupil diameter considered in our simulations. To scale the turbulence strength appropriately, the turbulence DL was placed at one-third of the distance between the first laser source and the collimating lens, corresponding to an effective pupil diameter of approximately 90 mm.

Under these conditions, the turbulence sequence was applied, and the resulting aberrations were recorded for both fields. Supplementary Figure 2 shows the RMS wavefront error per Zernike mode for the two fields. The average total RMS wavefront error was 0.43 $\lambda$ for the first (turbulent) field and 0.04 $\lambda$ for the second field, which remained unaffected and represents the residual system aberration.

## 4. Results and analysis

Table 4 summarizes the coupled power under three conditions: (1) no turbulence, (2) turbulence with the AO loop open (OFF, system aberration correction only), and (3) turbulence with the AO loop closed (ON). As expected, the coupling efficiency of the first field dropped nearly to zero under the presence of turbulence, while the second channel remained essentially unaffected.

Table 4: Average power coupled into the two fields with turbulence on field 1

|  | Correction OFF | Correction ON | No Turbulence |
|---|---|---|---|
| **Field 1** | 0.33μW | 1.47μW | 2.81μW |
| **Field 2** | 2.49μW | 1.87μW | 2.49μW |

When the AO loop was closed, turbulence was applied to Field 1 while recording both the optical fiber outputs and the PSFs from the far-field camera. The resulting coupling efficiencies for the three conditions—no turbulence, turbulence, and turbulence + MCAO correction—are shown in Figure 7. Fig. 8 shows the long exposure PSF for both spots in the corrected and uncorrected modes. Visualization 2 shows a movie of the spots without and with correction.

The average coupled powers, reported in Table 4, demonstrate that the system effectively corrected turbulence in the first field. While a modest reduction (~25%) was observed in the coupling efficiency of the second field, the first field exhibited an improvement of approximately 445%. The decrease in the second channel performance arises from the partial coupling between the two optical paths that can be both spatial and temporal. In this experimental implementation we estimated a rejection 3dB temporal bandwidth of about 12Hz due to the implementation of the code in Phyton. More advanced implementation in C++ on a real time operating systems could improve this up to 170Hz [17, 18] thus further improving the performance.

A comparison of the Zernike decompositions for the first field before and after correction (Supplementary Fig. 7) confirms that the system effectively corrected modes 1–5 (tip, tilt, defocus, and astigmatism), while higher-order aberrations remained largely uncorrected. Overall, the average wavefront RMS of Field 1 was reduced from 0.43 λ to 0.16 λ following correction.

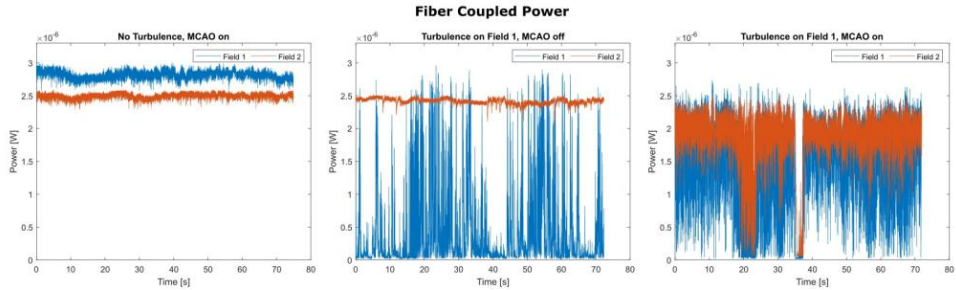

Fig. 7: coupled power for the three conditions: left, no turbulence, central turbulence ON on field 1, right turbulence ON field 1 and MCAO correction ON.

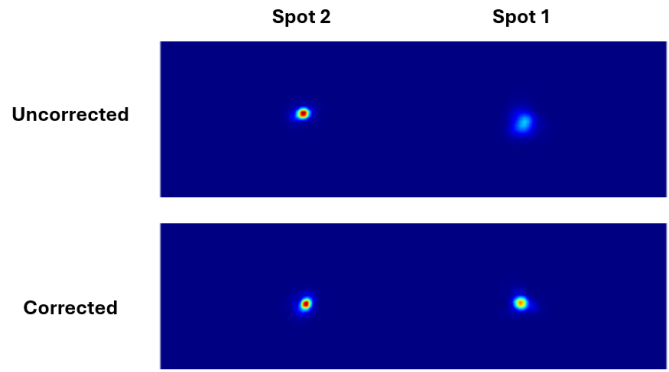

Fig. 8: Long exposure PSF images of both fields in the corrected and uncorrected mode. In Visualization 2 we show the far field spots acquired with the turbulence correction on and off.

## Conclusion

We have demonstrated, through both numerical simulations and laboratory experiments, that a Multi-Conjugate Adaptive Optics (MCAO) architecture based on multi-actuator deformable lenses is a practical and effective solution for wide-field correction of atmospheric turbulence. The transmissive nature of the DLs enables a compact optical layout, making this approach particularly suitable for portable free-space communication systems. Our R-MCAO implementation expanded the corrected field of view by a factor of three compared with conventional SCAO, while simultaneously stabilizing two spatially separated communication channels. The main current limitation is the control bandwidth, constrained to ~12 Hz by software latency (written in Phyton in this implementation). Future efforts will focus on accelerating the control pipeline, potentially through FPGA-based or compiled implementations on real time operating systems, to increase the rejection bandwidth in order to reduce the temporal error. The dual-fiber demonstration presented here opens the path toward systems employing bundles of single-mode fibers, which would allow substantial scaling of the communication capacity. In addition the use of this system can bring advantages in imaging systems in small telescopes.

## References

1. R. K. Tyson, *Introduction to Adaptive Optics*, 2nd ed. (SPIE Press, Bellingham, 2000).L. M. Rachbauer, D. Bouchet, U. Leonhardt, *et al*., J. Opt. Soc. Am. B **41**(9), 2122–2139 (2024).
2. M. J. Booth, *Adaptive optical microscopy: the ongoing quest for a perfect image*, Light: Science & Applications **3**, e165 (2014).
3. A. Roorda and D. R. Williams, "The arrangement of the three cone classes in the living human eye," *Nature* **397**, 520–522 (1999)
4. Bonora S, Pilar J, Lucianetti A, Mocek T. Design of deformable mirrors for high power lasers. *High Power Laser Science and Engineering*. 2016;4:e16. doi:10.1017/hpl.2016.14

5. Ragazzoni, R., Marchetti, E. & Rigaut, F. Modal tomography for adaptive optics. Astron. Astrophys. 342, L53±L56 (1999).
6. Ragazzoni, R., Marchetti, E. & Valente, G. Adaptive-optics corrections available for the whole sky. *Nature* **403**, 54–56 (2000). https://doi.org/10.1038/47425
7. M.Sohmen, et al. , "Optofluidic adaptive optics in multi-photon microscopy," Biomed. Opt. Express 14, 1562-1578 (2023)
8. T.Furieri, A.Bassi, S.Bonora, J. Biophotonics 2023, 16(12), e202300104. https://doi.org/10.1002/jbio.202300104
9. M.Quintavalla, et al "XSAO: an extremely small adaptive optics module for small-aperture telescopes with multiactuator adaptive lens," J. Astron. Telesc. Instrum. Syst. 6(2) 029004 (5 May 2020) https://doi.org/10.1117/1.JATIS.6.2.029004
10. M. Schiavon, et al. Multi-Actuator Lens Systems for Turbulence Correction in Free-Space Optical Communications. *Photonics* **2025**, *12*, 870. https://doi.org/10.3390/photonics12090870
11. L.Borsoi et al. High-efficiency free-space optical communication link with refractive adaptive optics 2025 Optics Express DOI: 10.1364/OE.577345
12. F.Santiago, et al, Underwater imaging with refractive adaptive optics modulators, *Opt. Eng.* **64**(12), 123101 (2025), doi: 10.1117/1.OE.64.12.123101.
13. J.Mocci et al. *"Analysis of Horizontal Atmospheric Turbulence by using a Shack-Hartmann Wavefront Sensor"*. In: AOIM XI, Murcia. 2018
14. S Bonora, et al, *"Wavefront correction and high-resolution in vivo OCT imaging with an objective integrated multi-actuator adaptive lens,"* Opt. Express 23, 21931-21941 (2015)
15. H. Verstraete, et al (2017). *Wavefront sensorless adaptive optics OCT with the DONE algorithm for in vivo human retinal imaging.* Biomedical optics express, 8(4), 2261-2275.
16. J.M. Bueno, et al, *"Wavefront correction in two-photon microscopy with a multi-actuator adaptive lens,"* Opt. Express 26, 14278-14287 (2018)
**17.** S.Bonora at al, Testing of a generic adaptive optics module (GAOM) for free-space quantum and classical communication over a 10 km link, 20 January 2026, Photonics West.
18. Vincent Deo, et al, The CACAO real-time computer for adaptive optics: updates, performance, and development plans, Proc. SPIE 13097, Adaptive Optics Systems IX, 130974O (28 August 2024); https://doi.org/10.1117/12.3020601

## Supplementary materials

### Deformable lenses

The adaptive lenses technology has already proven its capability both in ophthalmology, microscopy and astronomy [14][15][16]. The lens is produced by Dynamic Optics Srl and the lens design can be found in [14]. It is composed of two thin glass membranes upon each of which is glued a piezoelectric ring of actuators; once a voltage is applied the actuator behaves like a bimorph actuator. The number of actuators can be custom made, but standard DL have 18 or 32 actuators, this means that each glass membrane can have 9 or 16 actuators. The space between the two glass membranes is filled with PDMS (n=1.43). Each actuator can be independently controlled by applying a voltage that ranges between -75V to 125V by an High Voltage-USB driver. The influence functions of the lenses are shown in Supplementary Fig. 1.

The step time response of the DL (10% - 90%) is around 1.5ms with this driver. Such delay is totally acceptable in case of microscopy and it is usable even in astronomy/optical communication in case of low turbulence. The available clear apertures are 10, 16, 25mm, each one with 18 actuators, for this reason. The DL's diameters clearly impose a constrain on the position of the conjugated adaptive lenses, leaving the FoV angle as the only one that can be tuned.

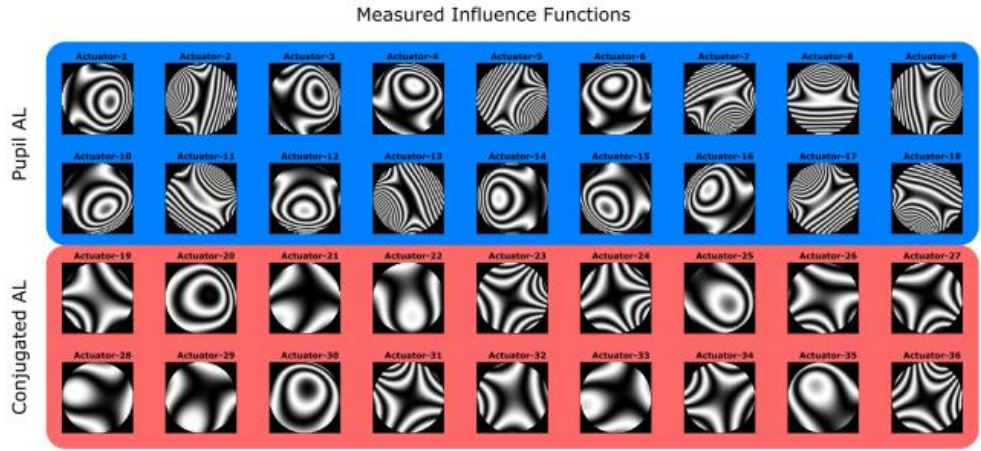

Supplementary Fig. 1: influence functions of the Pupil AL and of the Conjugated AL represented as interferograms at 633nm.

### Position of the deformable lenses

The formula to obtain the position x respect to the pupil plane of a given DL due to the geometrical constrain such as: FoV angle $\theta$, pupil DL clear aperture diameter $d$ and conjugated DL clear aperture diameter $D$ (see supplementary Fig. 2):

$$x = \frac{D - d}{2 tan \frac{\theta}{2}}$$

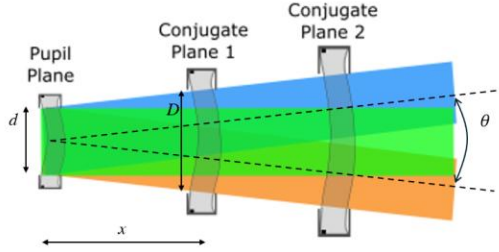

Supplementary Fig. 2: layout and position of the adaptive lenses.

## Modelling of the deformable lenses

Using a calibration setup equipped with a SH-WFS, we obtained an influence-function calibration for each deformable lens (DL). Each calibration can be represented as a 2D array in which every column contains 14,400 elements—corresponding to a $120 \times 120$ phase map generated by a single actuator. Thus, for the three DLs available, we obtained three influence matrices: IM1, IM2, and IM3.

As shown in Supplementary Fig. 2, for each beam corresponding to a specific position in the FoV, the conjugated DLs act only on a restricted portion of the full clear aperture. Based on this geometric consideration, once the desired FoV was defined, we extracted ("cut") the relevant portions of the influence functions of the conjugated DLs corresponding to all FoV beam positions.

To compensate for the reduced sampling introduced by this extraction, the resulting influence functions were up-sampled to recover the original $120 \times 120$ resolution. Supplementary Fig. 3 illustrates this procedure for a representative actuator and a specific FoV position.

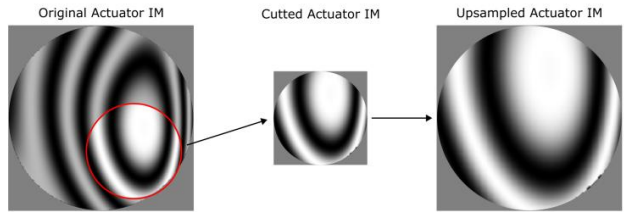

Supplementary Fig. 3: from left to right, a full influence function for an actuator of Conjugate DL. Center: the influence function "cut" to only the active region for a specific position in the FoV. Rigth: same influence function up-sampled to match the original resolution.

The IM is a map from actuator space to wavefront space. So in this framework $\varphi = (IM)\, c$ gives the wavefront information $\varphi$ due to the shape of the wavefront modulator obtained by the actuators poked as described by $c$.

A common procedure to deal with the inversion of *IM* it is the singular value decomposition (SVD) approach, i.e. the generalization of the eigen-decomposition of a square normal matrix to any n×m matrix.

Specifically, the singular value decomposition of an m×n real or complex matrix M is a factorization of the form $U\Sigma V^*$, where U is an m×m real or complex unitary matrix, $\Sigma$ is an

m×m rectangular diagonal matrix with non-negative real numbers on the diagonal, and V is an n×n real or complex unitary matrix. If M is real, U and $V^T=V^*$ are real orthonormal matrices.

In the column of U we can view the modes that our wavefront modulator is able to reproduce, while the singular values in the diagonal of Σ tells us the amplitude of that mode that can be reproduced.

We can now invert (IM) obtaining (CM) called Control Matrix CM:

$$\overline{\overline{CM}} = -V\Sigma_f^{-1}U^T$$

$$\overline{c} = \overline{\overline{CM}} \cdot \overline{\varphi} \qquad\qquad Eq.4$$

**MCAO control**

Starting from the previous configuration, we now introduce a second wavefront modulator conjugated to a different atmospheric layer, while keeping a single wavefront sensor conjugated to the pupil plane. The system now includes $n_1$ and $n_2$ actuators, and the control vector **c** is formed by concatenating the $n_1$ actuators of the first device with the $n_2$ actuators of the second device.

The procedure for constructing the Influence Matrix (IM) remains the same as before. Although the two devices are located at different conjugate planes, they are treated identically from a mathematical standpoint. Consequently, the IM takes the form:

$$\overline{\overline{IM}} = \begin{pmatrix} \overset{1st\ wavefront\ modulator}{\begin{bmatrix} W_{1,1} & \cdots & W_{1,n_1} \\ \vdots & & \vdots \\ W_{m,1} & \cdots & W_{m,n_1} \end{bmatrix}} & \overset{2nd\ wavefront\ modulator}{\begin{bmatrix} W_{1,n_1+1} & \cdots & W_{1,n_1+n_2} \\ \vdots & & \vdots \\ W_{m,n_1+1} & \cdots & W_{m,n_1+n_2} \end{bmatrix}} \end{pmatrix}$$

The layout highlights two block matrices, each inheriting the structure of Eq. 4. The first block contains, in its columns, the influence functions of the first wavefront modulator, while the second block contains those of the second wavefront modulator. The inversion process then follows the same procedure as before.

The only difference is that, in the SVD decomposition of the full Influence Matrix, it is not possible to directly identify the origin of each singular value and thus associate the modes with a specific wavefront modulator. However, the Influence Matrix can be split into its two constituent blocks and each block can be treated independently. This approach is particularly useful for visualization purposes—allowing the modes of each device to be distinguished—or for independently filtering out low–SNR modes. Once each block has been filtered, it can be substituted back into the original matrix. The Influence Matrix can then be inverted directly, without discarding any modes in the SVD of the already filtered matrix.

It is worth noting that, in this configuration, only a single point in the field of view is being measured. As a result, the use of multiple wavefront modulators offers no practical benefit and is presented here solely to support the understanding and development of the proposed framework.

**Two WFS's and one Wavefront Modulator**

In this case we have n1 actuators and m1 sampling points for the first WFS and m2 sampling points for the second WFS. Each WFS takes into consideration a different point of the field of view. In this case the IM takes this form:

$$\overline{\overline{IM}} = \begin{matrix} 1^{st} \text{ WFS} \\ 2^{nd} \text{ WFS} \end{matrix} \left( \begin{matrix} \begin{bmatrix} W_{1,1} & \cdots & W_{1,n} \\ \vdots & & \vdots \\ W_{m_1,1} & \cdots & W_{m_1,n} \end{bmatrix} \\ \begin{bmatrix} W_{m_1+1,1} & \cdots & W_{m_1+1,n} \\ \vdots & & \vdots \\ W_{m_1+m_2,1} & \cdots & W_{m_1+m_2,n} \end{bmatrix} \end{matrix} \right)$$

Even in this case it is a block matrix, where the top one is filled by the influence functions of the first field and the bottom one is filled by the influence functions of the second field.

The inversion process follows the same procedure as previously described. The Influence Matrix can be decomposed either as a whole, or—as mentioned earlier—by decomposing each block independently, filtering their singular values, and then reinserting the filtered blocks into their original positions within the Influence Matrix. The matrix can then be inverted directly using a pseudoinverse approach.

In this case, however, the correction stage is less straightforward. If the two portions of the field of view share the same aberration and the deformable device is conjugated to the pupil plane, a valid correction can be achieved simultaneously in both fields. Otherwise, the least-squares solution will determine the wavefront modulator shape that minimizes the measured gradients. Physically, this means that the controller attempts to compute an average correction between the two fields.

**Extension to two WFS's and two wavefront modulators**

At this point the extension takes this form:

$$\overline{\overline{IM}} = \begin{pmatrix} IM_{1,1} & IM_{1,2} \\ IM_{2,1} & IM_{2,2} \end{pmatrix} =$$

$$\left( \begin{matrix} \begin{bmatrix} W_{1,1} & \cdots & W_{1,n_1} \\ \vdots & & \vdots \\ W_{m_1,1} & \cdots & W_{m_1,n_1} \end{bmatrix} & \begin{bmatrix} W_{1,n_1+1} & \cdots & W_{1,n_1+n_2} \\ \vdots & & \vdots \\ W_{m_1,n_1+1} & \cdots & W_{m_1,n_1+n_2} \end{bmatrix} \\ \begin{bmatrix} W_{m_1+1,n_1+1} & \cdots & W_{m_1+1,n_1+n_2} \\ \vdots & & \vdots \\ W_{m_1+m_2,n_1+1} & \cdots & W_{m_1+m_2,n_1+n_2} \end{bmatrix} & \begin{bmatrix} W_{m_1+1,1} & \cdots & W_{m_1+1,n_1} \\ \vdots & & \vdots \\ W_{m_1+m_2,1} & \cdots & W_{m_1+m_2,n_1} \end{bmatrix} \end{matrix} \right)$$

All the considerations done on the single block can be extended here. In fact, acting on the single blocks we have the control on each field and each device.

## Correction zones

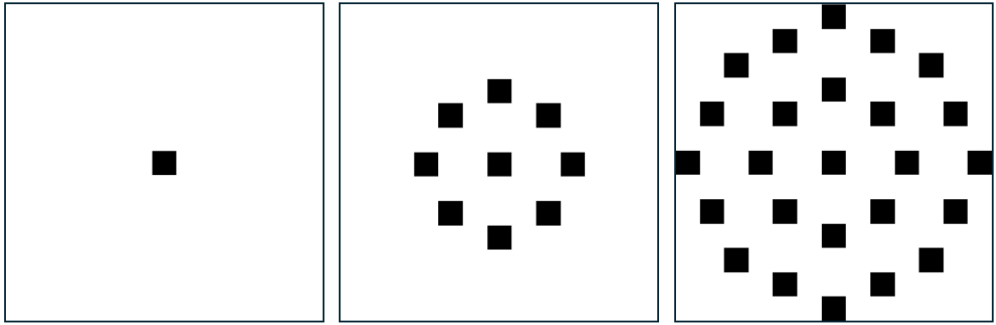

Supplementary Fig. 4: position of the corrected fields for the case of SCAO (left), small patch (central), large patch (right).

## Additional results

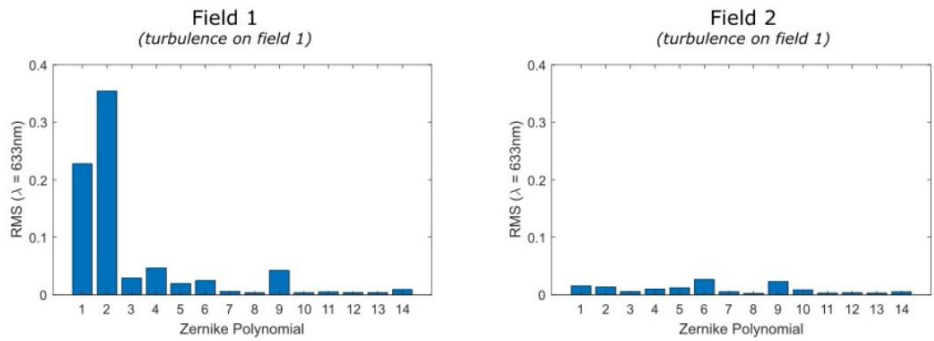

Supplementary Fig. 5: Zernike aberrations for both fields with turbulence generated on field 1.

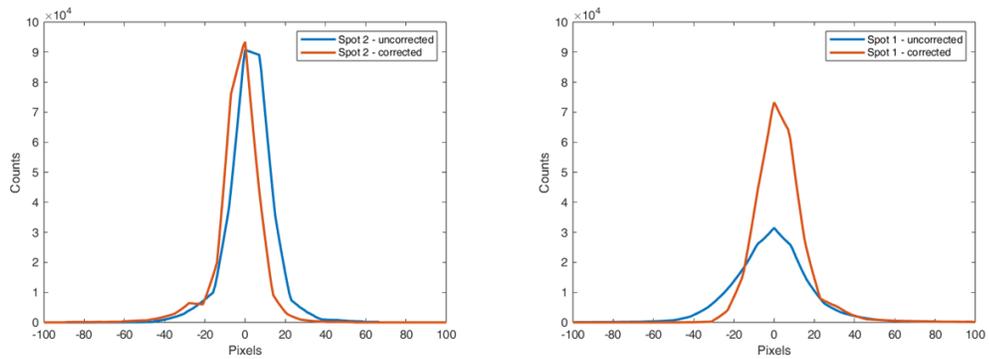

Supplementary Fig. 6: Cross section of long exposure PSF for both fields with turbulence generated on field 1.

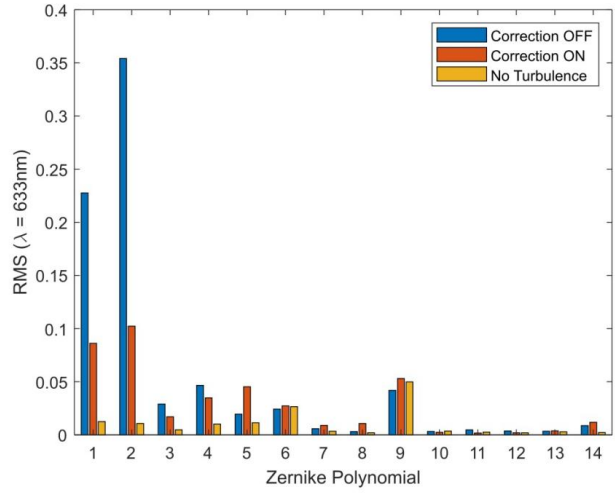

Supplementary Fig. 7: wavefront correction residuals

**Table 5.1:** Power measurement before and after the fiber injection without turbulence.

|  | Input Power [µW] | Output Power [µW] | coupling |
|---|---|---|---|
| **Field 1** | 5.46 | 2.81 | 51% |
| **Field 2** | 4.92 | 2.49 | 51% |